\documentclass[aps,prb,twocolumn,showpacs,superscriptaddress,10pt,longbibliography]{revtex4-2}
\usepackage{amssymb}
\usepackage{graphicx}
\usepackage[colorlinks = true,linkcolor = red,citecolor = blue,urlcolor=blue]{hyperref}
\usepackage[sort&compress]{natbib}
\usepackage{scalerel}
\usepackage[normalem]{ulem}
\usepackage[dvipsnames]{xcolor}
\usepackage{bm}
\usepackage{amsmath}
\usepackage{physics}
\usepackage{amsfonts}
\usepackage{mathtools}
\usepackage{dsfont}
\usepackage{float}
\usepackage{lmodern}
\usepackage[many]{tcolorbox}
\usepackage{empheq}
\usepackage{soul}
\usepackage[T1]{fontenc}
\usepackage{orcidlink}
\usepackage{qcircuit}

\begin{document}
\title{Dynamic Induction of Lattice Gauge Theories on a Quantum Computer}

\author{Bárbara Andrade${}^{\orcidlink{0000-0002-8268-3612}}$}
\email{bandradedoss@u-bordeaux.fr}
\affiliation{DIPC - Donostia International Physics Center, Paseo Manuel de Lardiz{\'a}bal 4, 20018 San Sebasti{\'a}n, Spain}
\affiliation{LaBRI - Laboratoire Bordelais de Recherche en Informatique, Universit{\'e} de Bordeaux, CNRS, 33405 Talence, France}
\author{Declan Millar${}^{\orcidlink{0000-0003-3713-8997}}$}
\affiliation{IBM Research, Hursley Park Road, Hursley, Winchester, SO21 2JN, United Kingdom}
\affiliation{School of Physics and Astronomy, University of Southampton, Southampton SO17 1BJ, UK}
\author{Lewis Anderson${}^{\orcidlink{0000-0003-0269-3237}}$}
\affiliation{IBM Research, Hursley Park Road, Hursley, Winchester, SO21 2JN, United Kingdom}
\author{Vincent R. Pascuzzi${}^{\orcidlink{0000-0003-3167-8773}}$}
\affiliation{IBM Research, 1101 Kitchawan Road, Yorktown Heights, NY 10598, USA}
\author{Maciej Lewenstein${}^{\orcidlink{0000-0002-0210-7800}}$}
\affiliation{ICFO - Institut de Ci{\`e}ncies Fot{\`o}niques, The Barcelona Institute of Science and Technology, 08860 Castelldefels, Spain}
\author{Ivano Tavernelli${}^{\orcidlink{0000-0001-5690-1981}}$}
\affiliation{IBM Research, S\"{a}umerstrasse 4, CH--8803 Rüschlikon, Switzerland}
\author{Jad C.~Halimeh${}^{\orcidlink{0000-0002-0659-7990}}$}
\affiliation{Department of Physics and Arnold Sommerfeld Center for Theoretical Physics (ASC), Ludwig Maximilian University of Munich, 80333 Munich, Germany}
\affiliation{Max Planck Institute of Quantum Optics, 85748 Garching, Germany}
\affiliation{Munich Center for Quantum Science and Technology (MCQST), 80799 Munich, Germany}
\affiliation{Department of Physics, College of Science, Kyung Hee University, Seoul 02447, Republic of Korea}
\author{Tobias Grass${}^{\orcidlink{0000-0002-8163-9353}}$}
\email{tobias.grass@dipc.org}
\affiliation{DIPC - Donostia International Physics Center, Paseo Manuel de Lardiz{\'a}bal 4, 20018 San Sebasti{\'a}n, Spain}
\affiliation{IKERBASQUE, Basque Foundation for Science, Plaza Euskadi 5, 48009 Bilbao, Spain}

\begin{abstract}
Gauge invariance is central to modern physics and underpins quantum simulations of lattice gauge theories (LGTs). Existing quantum simulation approaches employ Gauss's law either to energetically suppress gauge-violating processes in analog platforms or to detect and discard gauge-violating outcomes in digital devices. Here we introduce a third paradigm, in which Gauss's law is used to dynamically generate the gauge theory itself from a substantially simpler Hamiltonian. Starting from a readily programmable three-body XXX model, we employ experimentally efficient single-qubit U(1) gauge symmetry-generator terms that induce the dynamics of a U(1) LGT. We implement this approach using 101 qubits on a 156-qubit IBM quantum processor and observe real-time dynamics in quantitative agreement with the target LGT while reducing the entangling-gate depth per Trotter step by a factor of five compared with a direct implementation. Our results establish gauge protection as a resource for Hamiltonian engineering rather than merely symmetry preservation, opening a scalable resource-efficient route towards digital quantum simulations of increasingly complex gauge theories in higher spatial dimensions.
\end{abstract}

\maketitle

\textbf{Introduction.---} Gauge theories provide the fundamental language of modern physics, describing interactions ranging from quantum electrodynamics to quantum chromodynamics~\cite{Weinberg1995,Gattringer2009QuantumChromodynamicsLattice,Zee2003QuantumFieldTheory,Peskin1995}. Their lattice variants, lattice gauge theories (LGTs)~\cite{Rothe2012LatticeGaugeTheories}, have become a powerful framework for investigating salient phenomena in high-energy physics \cite{Kogut1975HamiltonianFormulationWilsons,Kogut1979AnIntroductionToLatticeGaugeTheory,Wilson1974,Wilson1977QuarksStringsLattice}, condensed matter \cite{wen2004quantum,Balents2010SpinLiquidsFrustrated,Savary2016QuantumSpinLiquids,Senthil20002GaugeTheory,Sedgewick2002FractionalizedPhase}, and quantum many-body dynamics \cite{Surace2020,Smith2017AbsenceOfErgodicity,Brenes2018ManyBodyLocalization,Budde2024QuantumManyBodyScars,Osborne2024QuantumManyBodyScarring,Cataldi2025DisorderFreeLocalizationFragmentation-1}. At the same time, LGTs constitute one of the most demanding targets for quantum simulation \cite{Bloch2008ManyBodyPhysics,Gross2017QuantumSimulations,Georgescu2014QuantumSimulation} because local gauge constraints generate highly non-trivial many-body dynamics while substantially increasing the complexity of experimental implementations \cite{Byrnes2006SimulatingLatticeGauge, Dalmonte2016LatticeGaugeTheory, Zohar2015QuantumSimulationsLattice,Mathis2020Towardscalablesimulations, Aidelsburger:2021mia, Zohar2021QuantumSimulationLattice, 
Barata2022MediumInducedJetBroadening,Klco2022StandardModelPhysics,Barata2023QuantumSimulationInMediumQCDJets,Barata2023RealTimeDynamicsofHyperonSpin, Bauer2023QuantumSimulationHighEnergy, Bauer2023QuantumSimulationFundamental,
DiMeglio2024QuantumComputingHighEnergy, Cheng2024EmergentGaugeTheory, Cohen2021QuantumAlgorithmsTransport,Barata2025ProbingCelestialEnergy, Lee2025QuantumComputingEnergy, Turro2024ClassicalQuantumComputing,Halimeh2023ColdatomQuantumSimulators,Bauer2025EfficientUseQuantum,Halimeh2025QuantumSimulationOutofequilibrium}.

\begin{figure*}[!t]
    \centering
    \includegraphics[width=0.95\linewidth]{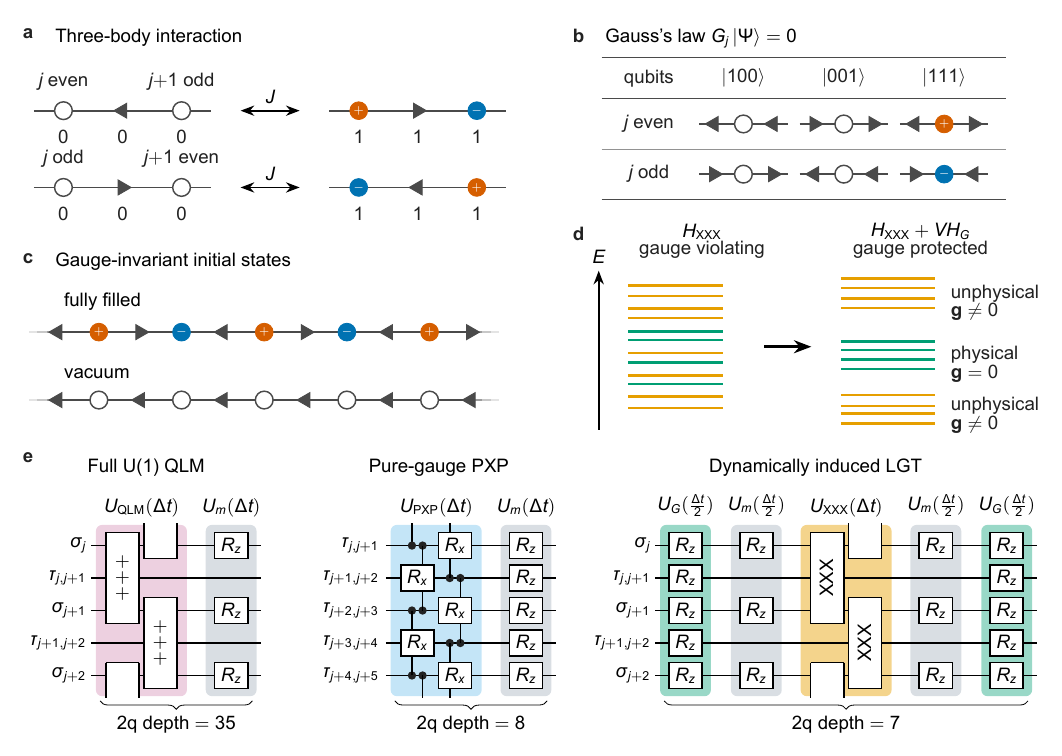}
    \caption{
        \label{fig:schematic}
        \textbf{Qubit encoding, gauge protection, and circuit realizations of the dynamically induced spin-$1/2$ $\mathrm{U}(1)$ quantum link Schwinger model.} \textbf{(a)} The gauge--matter interaction is a three-body term that creates a particle--antiparticle pair on neighboring matter sites and simultaneously flips the electric flux on the link between them, coupling $\ket{{+}{+}{+}}$ and $\ket{{-}{-}{-}}$ with strength $J$ in the rotated frame. \textbf{(b)} Gauss's law $G_j\ket{\mathrm{phys}}=0$ ties the matter charge at each site to the divergence of the electric field, leaving three gauge-invariant qubit triples per site parity. \textbf{(c)} The two initial states studied experimentally: the fully filled state and the bare vacuum. \textbf{(d)} In the spectrum of $H_{\mathrm{XXX}}$, the gauge-violating $\sigma^x\sigma^x\sigma^x$ interaction mixes the physical ($\mathbf{g}=0$, green) and unphysical ($\mathbf{g}\neq0$, orange) sectors; the single-body protection term $VH_G$ energetically separates them and confines the evolution to the physical, gauge-invariant subspace. \textbf{(e)} Trotterized circuit for a single step $\Delta t$ in the three implementations: the full $\mathrm{U}(1)$ QLM, the pure-gauge PXP model, and the dynamically induced LGT. Gauge protection reduces the two-qubit gate depth per step from $35$ to $7$.
}
\end{figure*}

The central challenge is enforcing Gauss's law. In analog quantum simulators, this is typically achieved by introducing energetic penalties that suppress gauge-violating processes, thereby restricting the dynamics to the physical Hilbert space \cite{Halimeh2022StabilizingGaugeTheories}. In digital quantum simulations, by contrast, gauge invariance is commonly exploited only after the computation, where measurements violating Gauss's law are discarded through post-selection or used for error mitigation \cite{Mildenberger2025}. Gauss's law can also be used in both digital and analog simulators to integrate out either the gauge \cite{Muschik2017} or the matter \cite{Surace2020} fields. While integrating out the gauge fields is fundamentally restricted to one spatial dimension, integrating out the matter fields applies to any spatial dimension but leads to highly nonlocal terms for $d\geq2$ spatial dimensions \cite{joshi2025efficientquditcircuitquench}. Although highly successful and have led to a suite of impressive quantum simulation experiments probing rich phenomena ranging from confinement and string breaking to thermalization and its avoidance \cite{Martinez2016RealtimeDynamicsLattice, Klco2018QuantumclassicalComputationSchwinger,Gorg2019RealizationDensitydependentPeierls, Schweizer2019FloquetApproachZ2, Mil2020ScalableRealizationLocal, Yang2020ObservationGaugeInvariance, Wang2022ObservationEmergent$mathbbZ_2$, Su2023ObservationManybodyScarring, Zhou2022ThermalizationDynamicsGauge, Wang2023InterrelatedThermalizationQuantum, Zhang2025ObservationMicroscopicConfinement, Zhu2024ProbingFalseVacuum, Ciavarella2021TrailheadQuantumSimulation, Ciavarella2022PreparationSU3Lattice, Ciavarella2023QuantumSimulationLattice-1, Ciavarella2024QuantumSimulationSU3, 
Gustafson2024PrimitiveQuantumGates, Gustafson2024PrimitiveQuantumGates-1, Lamm2024BlockEncodingsDiscrete, Farrell2023PreparationsQuantumSimulations-1, Farrell2023PreparationsQuantumSimulations, 
Farrell2024ScalableCircuitsPreparing,
Farrell2024QuantumSimulationsHadron, Li2024SequencyHierarchyTruncation, Zemlevskiy2025ScalableQuantumSimulations, Lewis2019QubitModelU1, Atas2021SU2HadronsQuantum, ARahman:2022tkr, Atas2023SimulatingOnedimensionalQuantum, Mendicelli2023RealTimeEvolution, Kavaki2024SquarePlaquettesTriamond, Than2024PhaseDiagramQuantum, Angelides:2023noe, Gyawali2025ObservationDisorderfreeLocalization,  
Mildenberger2025Confinement$$mathbbZ_2$$Lattice, Schuhmacher2025ObservationHadronScattering, Davoudi2025QuantumComputationHadron, Saner2025RealTimeObservationAharonovBohm, Xiang2025RealtimeScatteringFreezeout, Wang2025ObservationInelasticMeson,li2025frameworkquantumsimulationsenergyloss,mark2025observationballisticplasmamemory,froland2025simulatingfullygaugefixedsu2,Hudomal2025ErgodicityBreakingMeetsCriticality,hayata2026onsetthermalizationqdeformedsu2,Cochran2025VisualizingDynamicsCharges, Gonzalez-Cuadra2025ObservationStringBreaking, Crippa2024AnalysisConfinementString, De2024ObservationStringbreakingDynamics, Liu2024StringBreakingMechanism, Alexandrou:2025vaj,Cobos2025RealTimeDynamics2+1D,ilcic2026observationrobustcoherentnonabelian,chen2026thermalizationsu2latticegauge, Balaji:2025yua, Balaji:2025afl,xu2026observationglueballexcitationsstring, joshi2026observationgenuine21dstring,froland2026measuringnonstabilizernesssu2lattice}, these paradigms regard Gauss's law as a constraint that must be respected rather than as a resource for engineering quantum dynamics.

Here, we demonstrate a qualitatively different approach. Rather than implementing a LGT directly, we begin from a considerably simpler Hamiltonian that is naturally suited to digital quantum hardware but does not itself possess the desired gauge symmetry. We then use experimentally inexpensive single-qubit energy penalties proportional to the Gauss-law generators to dynamically confine the evolution to a physical sector with an emergent U(1) gauge symmetry. Within this constrained manifold, the effective evolution reproduces that of the target LGT, even though the underlying Hamiltonian is fundamentally different.

We experimentally realize this idea for the $1+1$D spin-$1/2$ U(1) quantum link formulation \cite{Chandrasekharan1997,Wiese2013UltracoldQuantumGases} of the Schwinger model \cite{schwingerGaugeInvarianceMass1962,Schwinger1962,Coleman1975,Coleman1976}. Starting from an easily programmable three-body XXX Hamiltonian, the dynamical gauge protection induces the desired gauge-theory dynamics while dramatically simplifying the quantum circuit. The resulting implementation requires only a single entangling layer per Trotter step, reducing the two-qubit gate depth by approximately a factor of five compared with a direct realization of the target Hamiltonian. Implemented on a chain of 101 qubits of IBM's 156-qubit \texttt{ibm\_basquecountry} processor, the induced gauge theory accurately reproduces the target real-time dynamics over experimentally accessible times.

Beyond providing an efficient route to digital quantum simulations of LGTs, our work establishes a broader paradigm in which symmetry constraints are elevated from passive conservation laws to active tools for Hamiltonian engineering. This perspective opens new opportunities for realizing complex many-body models from experimentally accessible interactions, particularly in settings where direct implementations are prohibitively demanding.

\textbf{Model.---} The target theory of our simulation is the spin-$1/2$ representation of the $\mathrm{U}(1)$ quantum link model (QLM) in $1+1$ dimensions \cite{Chandrasekharan1997,Wiese2013UltracoldQuantumGases,Kasper2017ImplementingQuantumElectrodynamics}. The staggered fermionic degrees of freedom are expressed by Pauli matrices $\sigma^i_{j}$ at site $j$ through a Jordan--Wigner transformation. The gauge bosons are represented by Pauli matrices $\tau^i_{j,\:j+1}$ acting on the gauge link $(j,j+1)$. A rotation $\mathcal{U} = \prod_j \sigma^x_{2j+1} \tau^x_{2j+1,\:2j+2}$ of every second fermion operator yields the target Hamiltonian
\begin{equation}
    H_{\rm{QLM}} = J \sum_{j=0}^{N_f-2} \left(\sigma^+_{j}\tau^+_{j,\:j+1}\sigma^+_{j+1} + \rm{h.c.}\right) - m \sum_{j=0}^{N_f-1} \sigma^z_j.
    \label{eq:QLM}
\end{equation}
Here, $N_f$ is the number of matter sites (even), $J$ is the strength of the three-body term which creates or annihilates matter/antimatter pairs on neighboring sites along with the corresponding change of gauge field configuration, and $m$ is the mass of the matter fields. The qubit implementation of this system is indicated in Fig.~\ref{fig:schematic}(a--c): The lattice consists of $L=2N_f+1$ qubits in total, with even qubits representing the staggered matter sites. On both particle and antiparticle sites, the state $\ket{0}$, defined by $\sigma^z \ket{0} = \ket{0}$, 
corresponds to an empty configuration. 
The state $\ket{1}$, defined by $\sigma^z \ket{1}=-\ket{1}$, corresponds to an occupied matter site.
The gauge field qubits in our encoding are in $\ket{1}$ if they surround an occupied matter site. If the staggered matter site in between is empty, our encoding  demands that neighboring gauge field qubits are in  different $\tau^z$ eigenstates, see Fig.~\ref{fig:schematic}(b). All physical configurations $\ket{\Psi}$  fulfill a $\mathrm{U}(1)$ Gauss's law, which at each matter site $j$ is expressed through $G_j\ket{\Psi}=0$, with the operator $G_j$ being defined as the difference between the electric field divergence and the charge density,
\begin{equation}
    G_j = \frac{(-1)^{j}}{2} \left( \sigma^z_j - \tau^z_{j-1,\:j} - \tau^z_{j,\:j+1} - 1 \right).
    \label{eq:Gauss}
\end{equation}
The gauge symmetry highly restricts the physical configurations, and we use the Gauss's law as an error-detection tool for our quantum computer experiment. Specifically, our experiment studies the dynamics of the QLM on initial states, such as a fully filled system or an empty system, see Fig.~\ref{fig:schematic}(c). To generate Trotterized QLM dynamics, we have pursued three different approaches, as illustrated in  Fig.~\ref{fig:schematic}(e):

(i) A first approach is based on a full QLM implementation. This requires three-body interactions between the configurations $\ket{000}$ and $\ket{111}$. A decomposition of this process into available $2$-qubit operations leads to a quantum circuit  with a $2$-qubit depth of $35$. This large depth severely limits the amount of Trotter steps that can be implemented with high fidelity. 

(ii) The PXP implementation uses Gauss's law to integrate out the matter fields, leaving a pure-gauge theory described by the PXP model \cite{Bernien2017, Turner2018Weak, Turner2018Quantum, Surace2020}
\begin{equation}
\begin{split}
    &H_{\rm{PXP}} = \sum_{j=0}^{N_f-2}P_{j-1,\:j}\tau^x_{j,\:j+1}P_{j+1,\:j+2}\\
    &\quad+ m\left(2\sum_{j=0}^{N_f-2}\tau_{j,\:j+1}^z + \tau_{-1,\:0}^z + \tau_{N_f-1,\:N_f}^z\right),
\end{split}
\end{equation}
where $P_{j,\:j+1}=\ket{1_{j,\:j+1}}\bra{1_{j,\:j+1}}$ is the projector onto the $\ket{1}$ state of gauge link $(j,j+1)$. A first order Trotter step under the PXP model has $2$-qubit depth of only $8$. Moreover, since the PXP dynamics does not create nearby excitations, we can use this gauge constraint for error-detection, although it is less restrictive than the gauge invariance of the full model. 

\begin{figure}[!t]
    \centering
    \includegraphics[width=\linewidth]{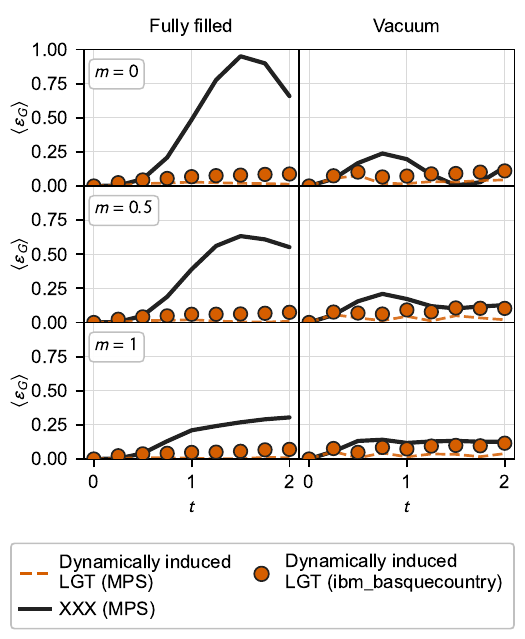}
    \caption{
        \label{fig:eGfull}
    \textbf{Gauge violation for a dynamically induced spin-$1/2$ $\mathrm{U}(1)$ quantum link Schwinger model.} Time evolution of the normalized average gauge violation $\expval{\varepsilon_G}$ following a quench from the fully-filled state (left) and vacuum state (right). Rows show staggered masses, $m=0$ (blue); $m=0.5$ (green); $m=1$ (orange). Solid lines show the evolution under $H_{\mathrm{XXX}}$ from Trotterized matrix-product-state (MPS) simulation; dashed lines give the MPS results for the dynamically induced gauge protection quantum circuits. Markers show \texttt{ibm\_basquecountry} hardware data for the dynamically induced LGT (circles). We set the protection strength to $V=4.5$ and use Trotter step size $\Delta t=0.25$.}
\end{figure}

(iii) A third implementation, with the smallest $2$-qubit circuit depth of only $7$, is based on a dynamically induced gauge symmetry.
Here we replace the three-qubit part of the QLM Hamiltonian by the Hamiltonian
\begin{equation}
    H_{\rm{XXX}} = J \sum_{j=0}^{N_f-2} \sigma^x_{j}\tau^x_{j,\:j+1}\sigma^x_{j+1} - m \sum_{j=0}^{N_f-1} \sigma^z_j.
\end{equation}
This Hamiltonian contains three-body terms that flip three neighboring qubits, no matter whether they are all aligned or not, facilitating the much shorter circuit depth, while maintaining all the processes which are present also in the QLM target theory. To reproduce the target theory, we need to suppress the flipping of unaligned spins, which, in first order, produce violations of the Gauss's law, Eq.~\eqref{eq:Gauss}. 
Therefore, we can achieve this suppression through a dynamical gauge protection scheme developed in Ref.~\cite{Halimeh2021}. Whereas that scheme, like energetic-penalty approaches more broadly~\cite{Halimeh2022StabilizingGaugeTheories}, was conceived to stabilize an intended gauge theory against gauge-breaking errors, we repurpose the same mechanism here to generate the target dynamics from a Hamiltonian that does not itself possess the Gauge symmetry. This scheme detunes all transitions that violate the Gauss's law just by applying single-qubit terms, thereby inducing a $\mathrm{U}(1)$ LGT which essentially matches the QLM model. Schematically, our approach is depicted in Fig.~\ref{fig:schematic}(d): While physical and unphysical configurations are energetically mixed in the spectrum of $H_{\rm{XXX}}$, they are energetically separated in the presence of a gauge protection term $VH_{\rm{G}}$. Importantly, this term is linear in $G_j$,
which facilitates its implementation only by single-qubit gates. Formally, the gauge protection Hamiltonian is given by $V H_{\rm{G}} = V \sum_{j=0}^{N_f-1} c_j \: G_j$, where $V$ is the gauge protection strength and $c_j$ are real numbers with $\abs{c_j}\leq1$. While the action of $H_{\rm{G}}$ on any physical state vanishes, for generic choices of $\{c_j\}$ transitions from physical to unphysical states will become off-resonant. Then, for a physical initial state the dynamics gets frozen to the physical sector through quantum Zeno effect. 
Different choices of gauge protection strength $V$ and sequences $\{c_j\}$ are investigated in Ref.~\cite{Halimeh2021}. For the Trotterized dynamics, there is an optimal window for $V$ depending on the Trotter time step size. In the experiment, we use $V=4.5$ for Trotter time steps $\Delta t=0.25$, and the simplest choice of sequence $c_j=(-1)^{j+1}$.
These choices highly suppress gauge violations across different lattice sizes. The structure of a Trotter step under the $H_{\rm{XXX}}+VH_{\rm{G}}$ Hamiltonian is represented in Fig.~\ref{fig:schematic}(e). The $\mathrm{XXX}$ blocks mutually commute, unlike the $+\!+\!+$ and PXP blocks. As a consequence, $U_{\mathrm{XXX}}$ carries no internal Trotter error and adds no further two-qubit depth in the second order Trotter evolution, which is not the case for the QLM and PXP.

\begin{figure}[!t]
    \centering
    \includegraphics[width=\linewidth]{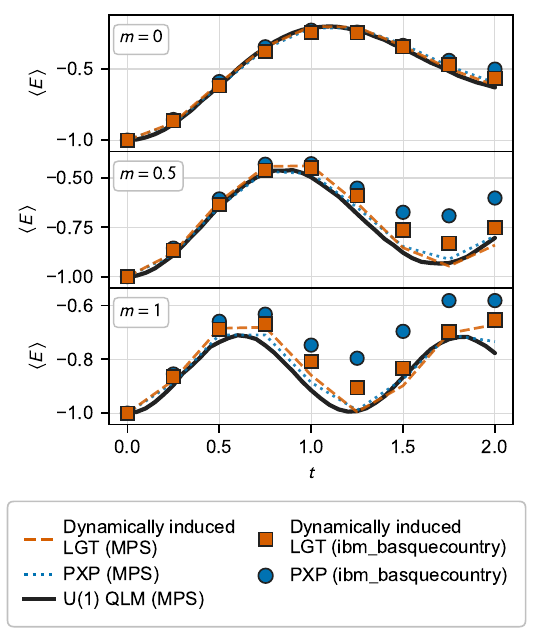}
    \caption{
        \label{fig:Efull}
    \textbf{Mean electric field for a dynamically induced spin-$1/2$ $\mathrm{U}(1)$ quantum link Schwinger model evolving from the fully-filled state.} Time evolution of the mean electric field $\langle E \rangle$ following a quench from the fully-occupied product state, where every matter site is occupied and gauge links are fixed by Gauss's law. Panels show staggered masses $m=0$ (blue); $m=0.5$ (green); $m=1$ (orange). Solid lines are the high-resolution target QLM trajectory from Trotterized matrix-product-state (MPS) simulation; dashed and dotted lines give the MPS results for the dynamically induced gauge protection and PXP circuits, respectively. Markers show \texttt{ibm\_basquecountry} hardware data for the PXP model (circles) and the dynamically induced LGT (squares). We set the protection strength to $V=4.5$ and use Trotter step size $\Delta t=0.25$.}
\end{figure}

Before turning to the quench dynamics, we first verify through MPS simulations that the gauge protection protocol effectively suppresses violations of Gauss's law during the evolution. Figure~\ref{fig:eGfull} demonstrates the action of the gauge protection mechanism by comparing the average gauge violation, quantified by $\expval*{\varepsilon_G}=\sum_j\expval*{G_j}^2/N_g$, where $N_g=N_f+1$ is the total number of gauge sites, during the evolution with ($V=4.5$, dashed red line) and without ($V=0$, solid black line) protection. In the absence of the gauge-protection term, the gauge violating three-body interaction rapidly drives the system out of the physical Hilbert space, leading to substantial accumulation of gauge violations for all initial states and masses. In contrast, introducing the gauge-protection term suppresses this growth by more than one order of magnitude throughout the evolution, in agreement with the continuous quantum Zeno dynamics picture discussed above.
Importantly, also the data from the quantum simulation measurements closely follows the MPS predictions of the gauge-protected system. This confirms that the gauge protection protocol remains effective under realistic experimental conditions. Clearly, hardware noise leads to residual gauge violations exceeding those of the ideal protected evolution, but the measured values remain substantially below those of the unprotected dynamics whenever gauge violations would otherwise become significant. Overall, the dynamics remains effectively confined to the physical gauge sector. This suppression is observed for both the vacuum and fully-filled initial states and across the full range of masses considered here, indicating the robustness of the protection scheme.

\begin{figure}[!t]
    \centering
    \includegraphics[width=\linewidth]{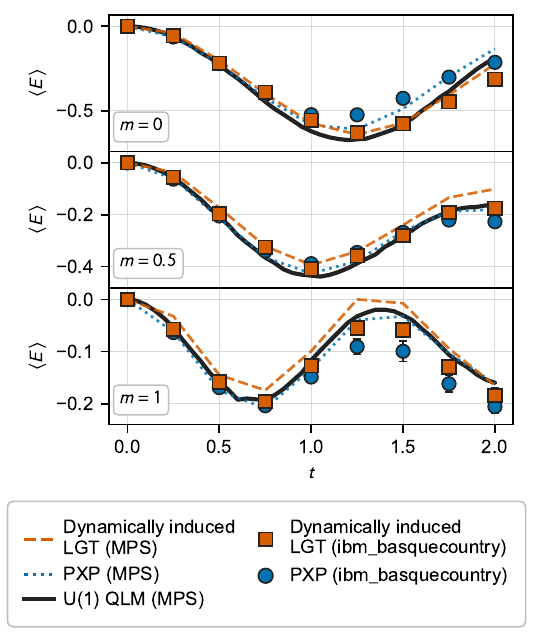}
    \caption{
     \label{fig:Evacuum}
    \textbf{Mean electric field for a dynamically induced spin-$1/2$ $\mathrm{U}(1)$ quantum link Schwinger model evolving from the vacuum.} Time evolution of the mean electric field $\langle E \rangle$ following a quench from the bare vacuum, where every matter site is empty and gauge links are fixed by Gauss's law. Panels show staggered masses $m=0$ (blue); $m=0.5$ (green); $m=1$ (orange). Solid lines are the high-resolution target QLM trajectory from Trotterized matrix-product-state (MPS) simulation; dashed and dotted lines give the MPS results for the dynamically induced gauge protection and PXP circuits, respectively. Markers show \texttt{ibm\_basquecountry} hardware data for the PXP model (circles) and the dynamically induced LGT (squares). We set the protection strength to $V=4.5$ and use Trotter step size $\Delta t=0.25$.}
\end{figure}

\textbf{Quench experiments.---} 
We now compare the different implementations by performing quench experiments on the quantum \texttt{ibm\_basquecountry} and using classical simulations on \texttt{Qiskit}'s MPS simulator as a benchmark~\cite{javadi2024quantum}. 
Since the implementation of the full QLM turns out to be too noisy, we concentrate on the PXP implementation and the dynamically protected XXX implementation. Specifically, we simulate the evolution of the vacuum and the fully-filled state under the dynamically induced LGT in second order Trotter formula and under the pure gauge PXP model in first order Trotter formula. In both schemes, we study systems with masses $m=0$, $0.5$ and $1$. Our lattice has $N_f=50$ matter sites, with the quantum simulations involving $101$ qubits under the dynamically induced LGT scheme and $51$ qubits under the PXP model. In all cases, the evolution spans $8$ Trotter steps of size $\Delta t=0.25$. 
For comparison we have also obtained a quasi-continuous evolution using \texttt{Qiskit}'s matrix-product states (MPS) classical simulator, where the Trotter steps size is only $\Delta t=0.05$ with a total of 40 first-order Trotter steps.

As a global figure of merit for our time evolution
we choose the mean electric field defined as $\expval*{E}=\langle \sum_{j=0}^{N_f-1} \tau_{j,\:j+1}^z \rangle /N_f$. In the empty vacuum configuration, it takes the value  $\expval*{E}=0$, whereas for a fully filled state the mean electric field is maximal in absolute value, $\expval*{E}=-1$.
Figure~\ref{fig:Efull} shows the time evolution of the mean electric field following a quench from the fully-filled state for three values of the mass. 
The kinetic term of the model drives pair annihilation processes which initially diminish the absolute value of the electric flux. Pair fluctuations then lead to oscillatory behavior of the electric flux.
Larger fermion mass progressively suppresses these processes, leading to smaller oscillation amplitudes. 
On the other hand, the oscillation periods become shorter for increased mass, in analogy to the effect of increased detuning in the case of single-qubit Rabi frequencies.
The three mass values therefore probe dynamical regimes ranging from large-amplitude excursions at $m=0$ to increasingly constrained dynamics at larger masses.
The analog dynamics, but starting from the empty state, is shown in Fig.~\ref{fig:Evacuum}. Now the kinetic term initially creates particle-antiparticle pairs, building up electric flux. As in the case of a fully filled initial state, increasing the mass suppresses the kinetic processes responsible for this dynamics, leading to smaller amplitudes of the electric field and shorter oscillation periods.

Both qualitatively and quantitatively, the above described dynamical behavior is not only featured by the QLM target theory (black lines in Figs.~\ref{fig:Efull} and \ref{fig:Evacuum}), but also by the theoretical benchmarks for Trotterized circuits simulating the dynamically induced implementation (red dashed lines) or the PXP implementation (blue dotted line). In the case of the PXP benchmark, deviations from the target line are only due to first-order Trotter errors, whereas in the case of the benchmark for the dynamically induced LGT, the deviations to the QLM target line are due to both second-order Trotter errors and the incomplete suppression of gauge violations, as measured in Fig.~\ref{fig:eGfull}. Importantly, the experimental data matches extremely well the corresponding theoretical benchmarks throughout the experimentally accessible time window of eight steps, both for the dynamically induced LGT circuit and for the PXP circuit. Small deviations occur only after several Trotter steps due to hardware errors. Notably, these errors become comparably larger in the PXP implementation, despite similar circuit depth and smaller number of qubits. This is due to the significantly smaller data post-selection, highlighting the importance of Gauss's law as an error mitigation strategy. 

\textbf{Discussion.---}
In this work, we report on the digital quantum simulation of the Schwinger model within the spin-$1/2$ U(1) quantum link formulation. This theory allows for a straightforward mapping of both matter and gauge degrees of freedom onto qubits. Using the \texttt{ibm\_basquecountry} device, we  have pursued three implementation strategies: While the circuit for a direct implementation is too deep for a faithful simulation, both the dynamically induced implementation with 101 qubits representing 50 matter sites and 51 gauge links, and the PXP implementation, which keeps only the 51 gauge qubits, yield quantitatively good results during up to eight Trotter steps. Despite the fact that the dynamically induced implementation only approximates the model, most of its data matches better the target theory, mainly due to a slightly shorter circuit (depth 7 instead of 8), and the large redundancy of the encoding providing a powerful error detection and post-selection strategy. Apart from outperforming other implementation schemes, the dynamically induced QLM generates important opportunities for future work. 
Our experiment demonstrates that single-qubit gates can enforce symmetries without the need to explicitly encode them in the circuits, opening an avenue to significantly simplified quantum circuits for complicated many-body interactions.

The success of our experiment also raises the question whether a dynamically induced gauge protection can also be used to suppress noise. To probe this, we have added the gauge protection layer of the circuit to the full QLM implementation, without observing a significant improvement of the data quality. This is in line with the intuition that the gauge protection schemes suppresses coherent transitions into non-physical states by making them strongly off-resonant, while not affecting these transitions when caused by incoherent noise \cite{Halimeh2020Fateoflatticegauge}.

One of the main advantages of the dynamically induced implementation is that it naturally extends to higher-dimensional LGTs, where alternative simplifications become increasingly impractical. In a $d$-dimensional U(1) QLM \cite{Zache2022}, the Gauss's law operators are generalized to
$ G_{\vec{n}} = \sum_j (S_{\vec{n},\vec{n}+\hat{e}_j}^z - S_{\vec{n}-\hat{e}_j,\vec{n}}^z) - \left[\frac{(-1)^{\vec{n}}-1}{2} + \psi_{\vec{n}}^{\dagger} \psi_{\vec{n}}\right]$, through the inclusion of the electric field divergence along every spatial direction, where $(-1)^{\vec{n}}\equiv(-1)^{\sum_i n_i}$ denotes the parity of the staggered site $\vec{n}$. As in the present work, the corresponding gauge protection Hamiltonian remains a linear combination of the Gauss's operators, $H_{\rm{G}}=\sum_{\vec{n}}c_{\vec{n}}G_{\vec{n}}$ \cite{Osborne2025Scale$2+1$DGaugeTheory}, and therefore requires only single-qubit rotations. Increasing the dimensionality merely increases the number of local terms entering each generator, without introducing additional entangling gates.
By contrast, eliminating the matter fields through Gauss's law rapidly becomes more involved in higher dimensions. While in $1+1$ dimensions each matter occupation is determined by the $2$ neighboring gauge links, in $2+1$ dimensions it depends on $4$ gauge links, so that the elementary gauge-matter interaction term $\psi_{\vec{n}} S_{\vec{n},\vec{n}+\hat{e}_j}^+ \psi_{\vec{n}+\hat{e}_j}$ maps onto an effective interaction involving $7$ qubits (with one shared link) \cite{joshi2025efficientquditcircuitquench}. Consequently, the locality and simplicity that make the PXP formulation attractive in one dimension are expected to deteriorate rapidly as the dimensionality increases, whereas the implementation cost of the dynamically induced approach is expected to increase much more gradually. Determining the optimal gauge-violating Hamiltonian whose protected dynamics reproduces the target theory in higher dimensions remains an important direction for future work.

\footnotesize
\textbf{Acknowledgments} 
B.A.~and T.G.~acknowledge support from the Department of Science, Universities, and Innovation of the Basque Government through the BasQ strategy (project EMISGALA) and the IKUR Strategy (project QT8 RyGaQuS), and from the Agencia Estatal de Investigación (AEI) through Proyectos de Generación de Conocimiento PID2022-142308NA-I00 (EXQUSMI). B.A.~acknowledges funding from the Maison du Quantique de Nouvelle-Aquitaine “HybQuant,” as part of the HQI initiative and France 2030, under French National Research Agency (ANR) Grant No. ANR-22-PNCQ-0002.
V.R.P.~was supported in part by the U.S.~Department of Energy, Office of Science, National Quantum Information Science Research Centers, Co-design Center for Quantum Advantage under Contract No.~DE-SC0012704. 
ICFO-QOT group acknowledges support from:
MCIN/AEI (PGC2018-0910.13039/501100011033,  CEX2019-000910-S/10.13039/501100011033, Plan National STAMEENA PID2022-139099NB, I00, project funded by MCIN/AEI/10.13039/501100011033 and by the “European Union NextGenerationEU/PRTR" (PRTR-C17.I1), FPI); Fundació Cellex; Fundació Mir-Puig;
Generalitat de Catalunya (European Social Fund FEDER and CERCA program;
Barcelona Supercomputing Center MareNostrum (FI-2023-3-0024);
EU funding  (EU Horizon 2020 FET-OPEN QU-ATTO, 101168628, HORIZON-CL4-2022-QUANTUM-02-SGA  PASQuanS2.1.  
J.C.H.~acknowledges funding by the Max Planck Society, the Deutsche Forschungsgemeinschaft (DFG, German Research Foundation) under Germany’s Excellence Strategy – EXC-2111 – 390814868, and the European Research Council (ERC) under the European Union’s Horizon Europe research and innovation program (Grant Agreement No.~101165667)—ERC Starting Grant QuSiGauge. This work is part of the Quantum Computing for High-Energy Physics (QC4HEP) working group.
\normalsize

\bibliography{Bibliography,references}

\clearpage
\onecolumngrid
\section*{Methods}

\subsection*{Quantum hardware and transpilation}

All quantum experiments were performed on IBM's superconducting quantum processor \texttt{ibm\_basquecountry} using the \texttt{SamplerV2} primitive. For each experiment, the logical qubits were mapped to a connected chain of physical qubits selected by maximizing the estimated layer fidelity computed from the reported two-qubit gate errors of the backend. Circuits were transpiled using \texttt{Qiskit}'s preset pass manager with optimization level $3$. Due to the stochastic nature of the transpiler, the pass manager was executed ten independent times for each circuit, and the transpiled circuit with the smallest depth was selected. During execution, gate twirling was enabled with 500 randomizations and automatic shot allocation across randomizations. Dynamical decoupling was enabled using the \texttt{XpXm} pulse sequence to suppress errors during qubit idle periods, and a repetition delay of $500\mu\mathrm{s}$ was used between consecutive circuit executions. Each data point was obtained from a total of $3\times10^5$ measurement shots, acquired over three independent runs of $10^5$ shots each performed on different days to reduce sensitivity to day-to-day calibration variations.

\subsection*{Circuit construction and Trotterization}

The quantum circuits implementing the Trotterized time evolution were constructed from elementary blocks corresponding to the local Hamiltonian terms. These blocks are assembled into the Trotter circuits shown schematically in Fig.~\ref{fig:schematic}(e) of the main text. The target QLM evolution operator $\exp{-i\Delta t\left(\sigma^+_j\sigma^+_{j+1}\sigma^+_{j+2}+\rm{h.c.}\right)}$ is implemented using the quantum circuit
\begin{center}
\scalebox{1.0}{
\Qcircuit @C=1.0em @R=0.2em @!R { \\
    \nghost{{q}_{0}:} & \lstick{{q}_{0}:} & \targ & \qw & \qw & \qw & \control\qw & \qw & \ctrl{1} & \qw & \qw & \qw & \targ & \qw & \push{}\\
    \nghost{{q}_{1}:} & \lstick{{q}_{1}:} & \ctrl{-1} & \ctrl{1} & \gate{\mathrm{R_X}(\Delta t)} & \gate{\mathrm{X}} & \ctrl{-1} & \ctrl{1} & \ctrl{1} & \gate{\mathrm{X}} & \gate{\mathrm{R_X}(\Delta t)} & \ctrl{1} & \ctrl{-1} & \qw & \push{,}\\
    \nghost{{q}_{2}:} & \lstick{{q}_{2}:} & \qw & \targ & \qw & \qw & \qw & \control\qw & \control\qw & \qw & \qw & \targ & \qw & \qw & \push{}\\
\\ }}
\end{center}
which exactly reproduces the corresponding three-body operator, while the mass term $\exp{-i\Delta t\left(-m \sigma^z_j\right)}$ is implemented with single-qubit RZ rotations. The first-order Trotter evolution quantum circuit for the QLM is obtained by sequentially applying this block to alternating triples of neighboring qubits followed by the mass layer. For the dynamically induced lattice gauge theory, the three-body interaction is replaced by the commuting $\mathrm{XXX}$ terms. The corresponding evolution operator, $\exp{-i\Delta t\left(\sigma^x_j\sigma^x_{j+1}\sigma^x_{j+2}\right)}$, is implemented using
\begin{center}
\scalebox{1.0}{
\Qcircuit @C=1.0em @R=0.2em @!R { \\
    \nghost{{q}_{0}:} & \lstick{{q}_{0}:} & \qw & \targ & \qw & \qw & \qw & \targ & \qw & \push{}\\
    \nghost{{q}_{1}:} & \lstick{{q}_{1}:} & \qw & \ctrl{-1} & \ctrl{1} & \gate{\mathrm{R_X}(2\Delta t)} & \ctrl{1} & \ctrl{-1} & \qw & \push{.}\\
    \nghost{{q}_{2}:} & \lstick{{q}_{2}:} & \qw & \qw & \targ & \qw & \targ & \qw & \qw & \push{}\\
\\ }}
\end{center}
This decomposition requires only four CNOT gates and a single RX rotation. Since all $\mathrm{XXX}$ terms mutually commute, the three-body evolution operator does not increase the two-qubit depth in the second order Trotter evolution. The gauge-protection and mass terms are implemented with single-qubit RZ rotations. The PXP evolution follows the first-order Trotterization and the interaction evolution operator $\exp{-i\Delta t\left(P_j\sigma^x_{j+1}P_{j+2}\right)}$ is implemented as
\begin{center}
\scalebox{1.0}{
\Qcircuit @C=1.0em @R=0.2em @!R { \\
    \nghost{{q}_{0}:} & \lstick{{q}_{0}:} & \qw & \qw & \ctrl{1} & \qw & \qw & \qw & \ctrl{1} & \qw & \qw & \qw & \qw & \push{}\\
    \nghost{{q}_{1}:} & \lstick{{q}_{1}:} & \qw & \gate{\mathrm{R_X}(\Delta t/2)} & \control\qw & \gate{\mathrm{R_X}(-\Delta t/2)} & \ctrl{1} & \gate{\mathrm{R_X}(\Delta t/2)} & \control\qw & \gate{\mathrm{R_X}(-\Delta t/2)} & \ctrl{1} & \qw & \qw & \push{.}\\
    \nghost{{q}_{2}:} & \lstick{{q}_{2}:} & \qw & \qw & \qw & \qw & \control\qw & \qw & \qw & \qw & \control\qw & \qw & \qw & \push{}\\
\\ }}
\end{center}
As neighboring PXP terms do not commute, the evolution is performed using the odd-even ordering shown in Fig.~\ref{fig:schematic}(e) of the main text. Furthermore, the CZ gates commute with the RZ rotations implementing the mass term. We use this freedom to reorder these commuting operations when constructing the quantum circuit, reducing the depth without modifying the implemented unitary.

\subsection*{Post-selection and error analysis}

All experimental results reported in this work were obtained from a total of $3\times10^5$ measurement shots for each data point. The measurements were acquired over three independent experimental runs performed on different days, with $10^5$ shots collected at each run. Post-selection was applied independently to each day's data before evaluating the observables. For the dynamically induced LGT (dLGT), measurement outcomes containing more than $2$ gauge violations were discarded. For the PXP model, measurement outcomes containing more than $2$ adjacent Rydberg excitations were discarded. The standard deviations and the fraction of shots retained after post-selection for both initial states and all parameters are reported in Tables~\ref{tab:std_post_full} and \ref{tab:std_post_vac}. The expectation values were obtained by first computing the mean value of the observable from each independent experimental run using the corresponding $10^5$ shots. The final expectation value was then calculated as the average of these three estimates, while the reported standard deviation corresponds to the standard deviation of the three measurements. This procedure provides an estimate of the experimental reproducibility across different days rather than the shot-noise uncertainty of a single experimental run.

As shown in Tables~\ref{tab:std_post_full} and \ref{tab:std_post_vac}, the fraction of accepted shots generally decreases with increasing Trotter step, particularly for the dynamically induced LGT experiments, where less than $1\%$ of the measurement outcomes are retained at the longest evolution times. Nevertheless, the independently acquired datasets exhibit excellent agreement, resulting in small standard deviation throughout all experiments. The largest standard deviation observed among all measurements is $0.021$, demonstrating the consistency of the independently acquired datasets, even when only a small fraction of the measurement shots is retained. The relatively low post-selection fraction in the $101$-qubit dynamically induced LGT experiment is expected because the system contains $50$ matter qubits, each of which can contribute to gauge violations. Consequently, the probability of observing a small number of gauge violations increases with system size. Importantly, however, the overall gauge violation remains small during the dynamics, as quantified by $\expval*{\varepsilon_G}=\sum_j\expval*{G_j}^2/N_g$. As shown in Fig.~\ref{fig:eGfull} of the main text, this quantity remains small over the entire evolution, demonstrating that the proposed scheme effectively suppresses gauge violations even though a small number of gauge violations are tolerated during post-selection.

\begin{table}[t]
\centering
\caption{Standard deviation of the mean electric field after post-selection for the fully-filled initial state. The percentage in parentheses indicates the fraction of shots retained after post-selection.}
\begin{tabular}{ccccccc}
\hline
& \multicolumn{2}{c}{$m=0$} & \multicolumn{2}{c}{$m=0.5$} & \multicolumn{2}{c}{$m=1$} \\
\cline{2-3}\cline{4-5}\cline{6-7}
Trotter step & dLGT & PXP & dLGT & PXP & dLGT & PXP \\
\hline
1 &
$0.004$ (27.25\%) & $0.003$ (99.90\%) &
$0.004$ (28.02\%) & $0.001$ (99.90\%) &
$0.005$ (27.63\%) & $0.002$ (99.89\%) \\

2 &
$0.003$ (8.81\%) & $0.003$ (97.89\%) &
$0.003$ (10.17\%) & $0.002$ (98.02\%) &
$0.004$ (11.15\%) & $0.003$ (98.20\%) \\

3 &
$0.002$ (3.77\%) & $0.003$ (89.09\%) &
$0.001$ (4.94\%) & $0.004$ (90.08\%) &
$0.003$ (6.02\%) & $0.003$ (93.41\%) \\

4 &
$0.001$ (1.45\%) & $0.005$ (74.85\%) &
$0.005$ (1.95\%) & $0.004$ (79.88\%) &
$0.009$ (2.48\%) & $0.003$ (89.16\%) \\

5 &
$0.002$ (0.70\%) & $0.004$ (61.80\%) &
$0.011$ (1.20\%) & $0.008$ (72.99\%) &
$0.011$ (1.50\%) & $0.002$ (86.17\%) \\

6 &
$0.004$ (0.40\%) & $0.003$ (52.19\%) &
$0.017$ (0.72\%) & $0.015$ (68.04\%) &
$0.010$ (0.77\%) & $0.005$ (81.25\%) \\

7 &
$0.012$ (0.20\%) & $0.004$ (45.34\%) &
$0.021$ (0.39\%) & $0.015$ (63.68\%) &
$0.013$ (0.31\%) & $0.005$ (74.51\%) \\

8 &
$0.013$ (0.12\%) & $0.005$ (40.36\%) &
$0.016$ (0.16\%) & $0.011$ (58.39\%) &
$0.007$ (0.17\%) & $0.004$ (67.11\%) \\
\hline
\end{tabular}
\label{tab:std_post_full}
\end{table}

\begin{table}[t]
\centering
\caption{Standard deviation of the mean electric field after post-selection for the vacuum initial state. The percentage in parentheses indicates the fraction of shots retained after post-selection.}
\begin{tabular}{ccccccc}
\hline
& \multicolumn{2}{c}{$m=0$} & \multicolumn{2}{c}{$m=0.5$} & \multicolumn{2}{c}{$m=1$} \\
\cline{2-3}\cline{4-5}\cline{6-7}
Trotter step & dLGT & PXP & dLGT & PXP & dLGT & PXP \\
\hline
1 &
$0.001$ (9.81\%) & $0.005$ (91.08\%) &
$0.002$ (8.41\%) & $0.004$ (91.11\%) &
$0.001$ (7.96\%) & $0.005$ (91.26\%) \\

2 &
$0.002$ (2.54\%) & $0.003$ (79.81\%) &
$0.001$ (5.75\%) & $0.003$ (80.64\%) &
$0.000$ (9.49\%) & $0.005$ (80.11\%) \\

3 &
$0.001$ (3.80\%) & $0.004$ (70.26\%) &
$0.000$ (3.98\%) & $0.003$ (69.00\%) &
$0.001$ (1.90\%) & $0.005$ (67.84\%) \\

4 &
$0.002$ (2.00\%) & $0.008$ (65.44\%) &
$0.003$ (0.93\%) & $0.005$ (59.90\%) &
$0.002$ (1.62\%) & $0.010$ (55.18\%) \\

5 &
$0.001$ (0.76\%) & $0.008$ (59.59\%) &
$0.006$ (0.76\%) & $0.002$ (50.76\%) &
$0.006$ (0.50\%) & $0.015$ (44.39\%) \\

6 &
$0.007$ (0.37\%) & $0.003$ (52.57\%) &
$0.013$ (0.17\%) & $0.006$ (43.42\%) &
$0.005$ (0.32\%) & $0.019$ (36.31\%) \\

7 &
$0.017$ (0.15\%) & $0.007$ (44.83\%) &
$0.005$ (0.14\%) & $0.011$ (35.55\%) &
$0.002$ (0.19\%) & $0.016$ (28.49\%) \\

8 &
$0.002$ (0.06\%) & $0.013$ (37.10\%) &
$0.008$ (0.09\%) & $0.011$ (29.81\%) &
$0.015$ (0.06\%) & $0.012$ (23.44\%) \\
\hline
\end{tabular}
\label{tab:std_post_vac}
\end{table}

\subsection*{Matrix-product-state simulations}

Classical reference simulations were performed with matrix-product-states (MPS) method implemented using \texttt{Qiskit Aer}. For the system sizes considered in this work, we employed a modified version of the \texttt{AerSimulator} in which the maximum number of qubits supported by the MPS backend was increased from the default limit of $63$ to $156$. This modification affects only the backend configuration specifying the maximum accessible system size and does not alter the MPS simulation algorithm.

We used this simulator to obtain a high-resolution classical reference for the target QLM dynamics and to benchmark the finite-step quantum circuits implemented experimentally for the dynamically induced LGT and the pure-gauge PXP model. The target QLM evolution was simulated using a first-order Trotter formula with a time step of $\Delta t=0.05$ for $40$ Trotter steps, corresponding to a total evolution time of $t=2$. The experimentally implemented dynamically induced LGT and PXP dynamics consisted of $8$ Trotter steps with $\Delta t=0.25$, which were classically simulated using the same product-formula construction and time discretization as their corresponding quantum circuits.

The MPS truncation threshold was set to $10^{-12}$ for all simulations reported in this work. To assess convergence with respect to the MPS truncation threshold, we additionally performed simulations with thresholds of $10^{-8}$ and $10^{-10}$. The resulting dynamics exhibited negligible changes as the threshold was decreased, indicating that the observables considered here are well converged at the chosen value of $10^{-12}$.

To reproduce the measurements and post-selection used in the quantum computer experiments, the MPS simulated circuits were sampled using \texttt{Qiskit Aer}'s \texttt{Sampler V2} with $10^5$ shots per circuit. The resulting bit strings were processed using the same post-selection procedure as the experimental data: For the dynamically induced LGT, samples violating the Gauss's law on more than $2$ matter sites were discarded, whereas for the PXP implementation, samples violating the Rydberg blockade constraint on more than $2$ sites were discarded. Observables were then evaluated from the post-selected samples, enabling a direct comparison between the ideal simulated circuits and their experimental implementations.

\end{document}